# Diffusion of Be in gallium nitride: Experiment and modelling.

Rafał Jakieła[1], Kacper Sierakowski[2], Tomasz Sochacki[2], Małgorzata Iwińska[2], Michał Fijalkowski[2], Adam Barcz[1,4], and Michał Boćkowski[2,3]

[1] *Institute of Physics, Polish Academy of Sciences, Aleja Lotnikow 32/46, PL-02668, Warsaw, Poland*
[2] *Institute of High Pressure Physics, Polish Academy of Sciences, Sokolowska 29/37, 01-142, Warsaw, Poland*
[3] *Center for Integrated Research of Future Electronics, Institute of Materials and Systems for Sustainability, Nagoya University, C3-1 Furo-cho, Chikusa-ku, Nagoya 464-8603, Japan*
[4] *Institute of Electron Technology, Aleja Lotnikow 32/46, PL-02668, Warsaw, Poland*

Diffusion mechanism of beryllium in gallium nitride was investigated by analyzing temperature-dependent diffusion profiles from an infinite source. Beryllium atoms were implanted into a high structural quality gallium nitride layer crystallized by halide vapor phase epitaxy on an ammonothermal gallium nitride substrate. Post-implantation annealing was performed at different temperatures, between 1000°C and 1400°C, under high nitrogen pressure. Beryllium profiles were analyzed in the as-implanted and annealed samples by secondary ion mass spectrometry. It was shown that the diffusion of the dopant results from the combination of two mechanisms: rapid interstitial and slow interstitial-substitutional diffusion. The pre-exponential factor as well as activation energy for both diffusion paths were determined. Moreover, from the characteristic features of beryllium depth profiles, the formation energies of gallium vacancy and beryllium in interstitial position were calculated and compared to the theoretical values.

# 1. Introduction

It is well known that gallium nitride (GaN) and its alloys with indium and aluminum constitute a great base for optoelectronic as well as high power and high frequency electronic devices. Today, the nitride semiconductors have a status of the second technologically most important electronic materials after silicon (Si) [1]. All nitride-based devices are prepared by epitaxy. On the other side, it is well known that ion implantation is one of the basic tools for semiconductor device fabrication. The implantation process has been commonly applied for controlling the selective area doping of both n- and p-type regions, which allows reducing device size and controlling the electric field configuration in devices. In the case of GaN, high n-type carrier concentration and conductivity have already been demonstrated by using relatively-highly ion dosage [2,3]. High p-type conductivity after ion implantation still remains a challenge. Recently, very effective activation by ultra-high pressure annealing (UHPA) of magnesium (Mg)-implanted p-type GaN has been announced [4]. Investigation of Mg diffusion during the UHPA process also started [5]. Next to Mg, beryllium (Be) is one of the most promising acceptor dopants for GaN. Since a Be ion has a smaller mass than a Mg ion, Be ions allow reducing the implantation damage in comparison to Mg ones for a given implantation depth. The theoretical value of Be ionization energy is equal to 550 meV when Be resides on a Ga lattice site [6]. However, the experimental values of ionization energy were lower. Sanchez et al. reported 90 meV for MBE-grown and Be-doped GaN layers on Si wafers [7]. Nakano et al. experimentally reported p-type conduction of Be-implanted GaN with and without introduction of oxygen, where the ionization energies were 163 and 240 meV, respectively [8]. These values are comparable to that for an Mg on a Ga site. In the pioneer work by Nakano et al., the hole concentration at room temperature was less than 1% compared with introduced Be ions even after optimizing the annealing temperature. This can be explained by a considerable part of Be atoms sitting in interstitial positions after annealing and acting as donors compensating the acceptor dopants [9]. Such occupation of unexpected sites can be related to the diffusion process of Be atoms during annealing. Wang et al. implanted Be ions into an MBE GaN layer grown on sapphire [10]. They activated the Be ions by rapid thermal annealing and pulsed laser annealing. The latter method allowed to obtain hole sheet concentration of $2.56 \times 10^{13}$ cm$^{-3}$. The activation energy was 135 meV.

There is no extensive research reported concerning the Be diffusion in GaN. Additionally, post-implanted annealing at relatively low temperature (up to 1100°C) did not allow to properly observe the diffusion process [11,12,13,14]. Moreover, the cited results were obtained for highly defected heteroepitaxial GaN structures. It is well known that threading dislocations of high density can also act as diffusion passes. Diffusion coefficients of Be in GaN were reported by Koskelo et al. [15]. The coefficients were, however, calculated for samples annealed at 850ºC and 950ºC and co-doped with Li. Theoretical studies of Be diffusion in GaN were presented by Miceli et al. [16]. The possible diffusion paths of Be parallel and perpendicular to the c-axis were analyzed. In the first case it was shown that expected positions of Be atoms in adjacent octahedral volumes and the lowest energy path between these positions.

In this paper we describe a thorough investigation of Be diffusion in GaN. Concentration-dependent diffusion from an infinite source was analyzed. Halide vapor phase epitaxy (HVPE) layers, grown on ammonothermal GaN (Am-GaN) seeds, were used as samples for implantation of Be ions. Such HVPE-GaN layers exhibit high structural quality. The threading dislocation density (TDD) is at the level of $5\times10^4$ cm$^{-2}$ [15]. This value is at least two orders of magnitude lower than in case of heteroepitaxially-grown GaN layers. Such low TDD value allows to assume that dislocations will not play a significant role in the diffusion process of implanted ions. After implantation runs the samples were annealed by UHPA technology at temperature between 1000ºC and 1400ºC. This allowed to completely remove the implantation damage and started the diffusion processes in GaN. Secondary ion mass spectrometry (SIMS) Be in-depth profiles were determined. Two mechanisms of Be diffusion were revealed: rapid and slow. We suggest and discuss that two diffusion behaviors originate from interstitial and interstitial-substitutional paths of Be atoms. An infinite source model of diffusion was applied and the pre-exponential factor and diffusion activation energy for the rapid mechanism were established. In turn, the Boltzmann-Matano method was used to determine the parameters for the slower diffusion mechanism. Assuming the existence of gallium vacancies ($V_{Ga}$) in HVPE-GaN and basing on the SIMS Be depth profiles, $V_{Ga}$ concentration and Be solubility were determined for each annealing temperature. It allowed to calculate the energy of formation of the $V_{Ga}$ defect as well as Be at interstitial position.

## 2. Experimental details

An unintentionally doped 400-µm-thick HVPE-GaN layer deposited on an n-type 1-inch Am-GaN substrate was used for preparing samples for ion implantation. The HVPE-GaN was of high structural quality with etch pit density (EPD; well correlated with TDD) of the order of $5\times10^4$ cm$^{-2}$. The bowing radius of crystallographic planes was larger than 10 m. The free carrier concentration of HVPE-GaN layer was lower than $5\times10^{16}$ cm$^{-3}$. The HVPE-GaN/Am-GaN couple was sliced into the 5 mm × 5 mm-square pieces. All (0001) HVPE-GaN/Am-GaN surfaces of the samples were prepared to the epi-ready state by lapping, polishing and chemo-mechanical polishing (CMP). Details of ammonothermal, HVPE, and wafer fabrication technologies are described elsewhere [17,18].

Implantation of Be ions was performed into the described above HVPE-GaN layers. Two conditions of Be ions implantation were employed at room temperature without the use of a through film. The first condition was a dose of $5\times10^{15}$ cm$^{-2}$ at the energy of 150 keV (Implantation 1). The second one was a dose of $2.9\times10^{15}$ cm$^{-2}$ at the energy of 200 keV (Implantation 2).

Two series of UHPA processes were performed as post-implantation annealing. The UHPA technology comes directly from the high nitrogen pressure solution (HNPS) growth method well described elsewhere [19,20]. The first UHPA was performed for the samples from Implantation 1. They were annealed at 1000°C, 1200°C, and 1400°C under nitrogen pressure of 10 MPa, 200 MPa, and 800 MPa, respectively, and for a constant time of 15 minutes. Samples from Implantation 2 were annealed at constant nitrogen pressure of 1 GPa but at five different temperatures in the range of 1200 – 1400°C and two different times 15 and 30 minutes. The applied nitrogen pressures were much higher than the equilibrium nitrogen partial pressures for the respective temperatures in the phase diagram of a GaN-Ga-N$_2$ system [19], allowing annealing in the GaN solid phase without surface decomposition.

Before and after ion implantation, as well as after the UPHA process the samples were characterized by 2theta-omega scan of X-ray diffraction and SIMS. The Be depth profiles and other possible impurities were examined by SIMS technique using a CAMECA IMS6F microanalyzer. Molecular oxygen ions (O$_2^+$) were used as primary ions with the beam energy and the current of 8 keV and 800 nA, respectively. The size of the raster was about 150 µm ×150 µm and the secondary ions were collected from a central region of approximately 60 µm in diameter. Magnitudes of Be concentrations were derived from the intensity of Be$^+$ species with the matrix signal N$^+$ taken as a reference. A non-annealed Be-implanted GaN sample served as a calibration standard. The inaccuracy of all values calculated from SIMS depth profiles was determined using multiple measurements. C-planes of all samples analyzed by SIMS were covered with gold using sputter deposition. Without the sputtered metal, the implanted samples were non-conductive (highly resistive) and the SIMS measurements could not be carried out.

## 3. Results

The HVPE-GaN used for further implantation contains oxygen and silicon at the concentration level lower than $10^{17}$ at/cm$^3$ as measured by SIMS. It should be noted that all elements apart from Be, especially atmospheric impurities like hydrogen and carbon, were below the SIMS background level.

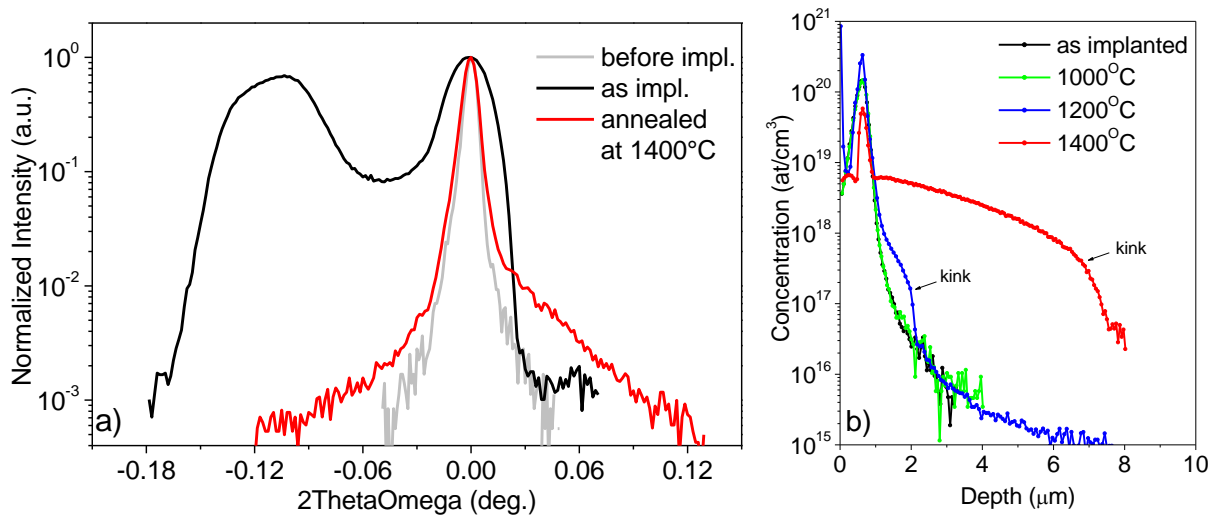

Fig. 1 a) X-ray 2theta-omega scan for samples before and after ion implantation (Implantation 1), as well as after first set of the UPHA process at 1400 °C; b) Be depth profiles for the samples annealed at temperatures of 1000°C, 1200°C and 1400°C under nitrogen pressure of 10 MPa, 200 MPa and 800 MPa, respectively, for 15 min.

Figure 1a presents X-ray 2theta omega scans for samples before and after ion implantation, as well as after the first set of the UPHA process at 1400°C. It is clearly seen that after Be implantation the sample lost its high structural quality. For the as-implanted sample, two broad peaks were visible in the scan. The peak which appeared in the lower angle was due to the extended lattice constant along the c-axis in the implanted layer compared to that in the host material. The peak broadening indicates the crystal lattice distortion induced by the implantation damage. Annealing at 1000°C allowed to reduce the width of the peaks (not shown in Fig. 1a). Higher temperatures (1200°C and 1400°C) restored the result of pre-implantation X-ray measurement. Only one narrow peak was detected. The spectrum of the sample annealed at 1400° is presented. A detailed analysis of the X-ray spectra of implanted and annealed samples (i.e. the broad shoulder in the high-angle side for samples annealed at 1400°C) will be analyzed in the future. Figure 1b shows depth profiles of Be atoms for the first set of annealed samples compared to a sample with no temperature treatment. Only the Be profile is presented. No significant changes in concentrations of other elements were observed. An increase of hydrogen concentration to the level of $10^{18}$ at/cm$^3$ after annealing at the temperature of 1400ºC was detected but it was not analyzed in this paper. According to Nartia et al. [5] unintentional moisture may exist in the UHPA system because of the difficulty of achieving a perfect purge. The moisture can become a hydrogen source under high pressure and temperature. The data presented in Fig. 1b indicate that: (i) the detection limit of Be is at the level of $10^{15}$ at/cm$^3$. (ii) annealing at 1000°C does not change the Be depth profile; (iii) annealing at 1200°C and 1400 °C leads to a change in the Be depth profile; Be reaches the depth of 8 μm at the 1400ºC annealing; (iv) both diffusion profiles at 1200°C and 1400°C exhibit a characteristic kink (as marked in Fig. 1b) at the concentration level of ~2×10$^{17}$ at/cm$^3$ and ~4×10$^{17}$ at/cm$^3$, respectively; (v) in the sample annealed at 1400°C the Be reservoir remains at the depth of maximum concentration of the as-implanted sample; it indicates that the top of the layer, around 1 μm from the surface, can be regarded as an infinite source of Be dopant for all annealing conditions.

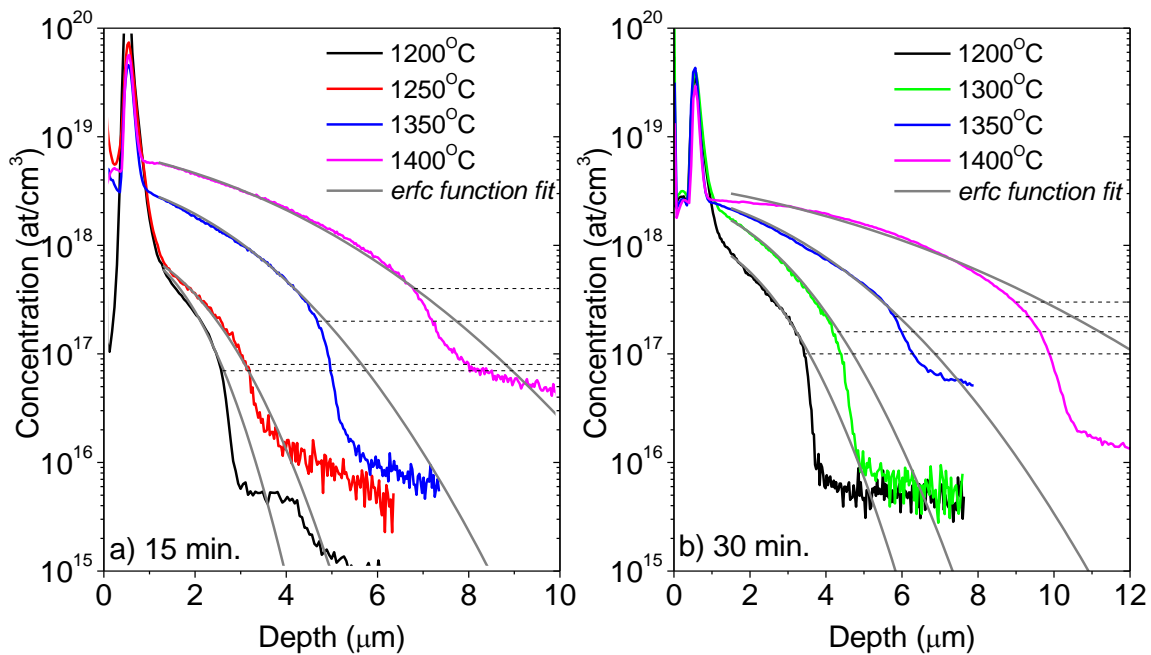

Fig. 2 Be depth profiles for samples annealed in the temperature range 1200°C – 1400°C for: a) 15 min, b) 30 min.; gray curves are *erfc* function fitting; black dotted lines indicate the concentration level of the kink in profile.

Figure 2 shows the Be depth profiles for the samples of the second series of annealing. If an infinite source of a species is used, the diffusion profiles (concentration $C$ of the examined species) should be described by a complementary error function – *erfc*:

$$C(x,t) = C_S erfc\left(\frac{x}{\sqrt{4Dt}}\right) \quad (1)$$

where: $C_S$ is the maximum concentration of the diffused species, which corresponds to the surface concentration in the case of the infinite source experiment; $D$ is the diffusion coefficient; $t$ is the time of annealing; and $x$ measures the depth from the source of species.

The fitting curves based on *erfc* functions are also presented in Fig. 2, where $C_S$ and $D$ were used as fitting parameters. The magnitudes of the diffusion coefficients obtained from fitting of Eq. (1) change substantially from $4 \times 10^{-12}$ cm²/s to $6 \times 10^{-11}$ cm²/s as the annealing temperature varies from 1200°C to 1400°C. However, it is clearly seen that the *erfc* relation is well fitted to the SIMS data only in the upper parts of the profiles (above the characteristic kink). Therefore, the calculated diffusion coefficients are valid for higher Be concentration. As in the case of the first set of samples, a drop in each Be depth profile is visible at a certain Be concentration, different for each sample. Additionally, the concentration level at which the drop occurs becomes higher with increasing the temperature. Profile deviation from *erfc* fitting indicates a change in the diffusion coefficient.

For systems in which the diffusion coefficient is a function of concentration, the method described by Matano [21] must be used. In such case the standard Fick's law equation is transformed into:

$$\frac{\partial C}{\partial t} = \frac{\partial}{\partial x}D\left(\frac{\partial C}{\partial x}\right) \quad (2)$$

where: $C$ is the atom concentration, $D$ is diffusion coefficient, and $t$ and $x$ are time and depth variables, respectively. When boundary conditions are provided, the following variable can be used: $\eta = x / t^{1/2}$. This results in the dependence of concentration only on $\eta$ instead of $x$ and $t$. Then, the equation can be integrated with respect to $\eta$ between $C = 0$ and $C = C_1$, where $C_1$ is some specific value of concentration. Since the analyzed experimental profile is plotted for a specific diffusion time, $t$ can be treated as constant when $\eta$ is replaced by $x$ and $t$. For $C$ equal

0 $dC/dx$ is also equal 0. Therefore, the final equation for the diffusion coefficient is the following:

$$D(C^1) = -\frac{1}{2t}\left(\frac{dx}{dC}\right)\bigg|_{C^1} \int_0^{C_1} x dC \qquad (3)$$

This way, the diffusion coefficient for a particular concentration $C^1$ can be derived from a dopant depth profile by transforming the plot from $C(x)$ to $x(C)$ function and integrating. Beryllium depth profiles transformed using equation (3) are presented in Figs 3a and 3b for 15-minute and 30-minute annealing, respectively. The Boltzmann–Matano method allowed to determine the diffusion coefficients for lower concentrations of Be, where there is no matching between the *erfc* fit and the SIMS data. The values of the coefficients are indicated by arrows in Fig. 3.

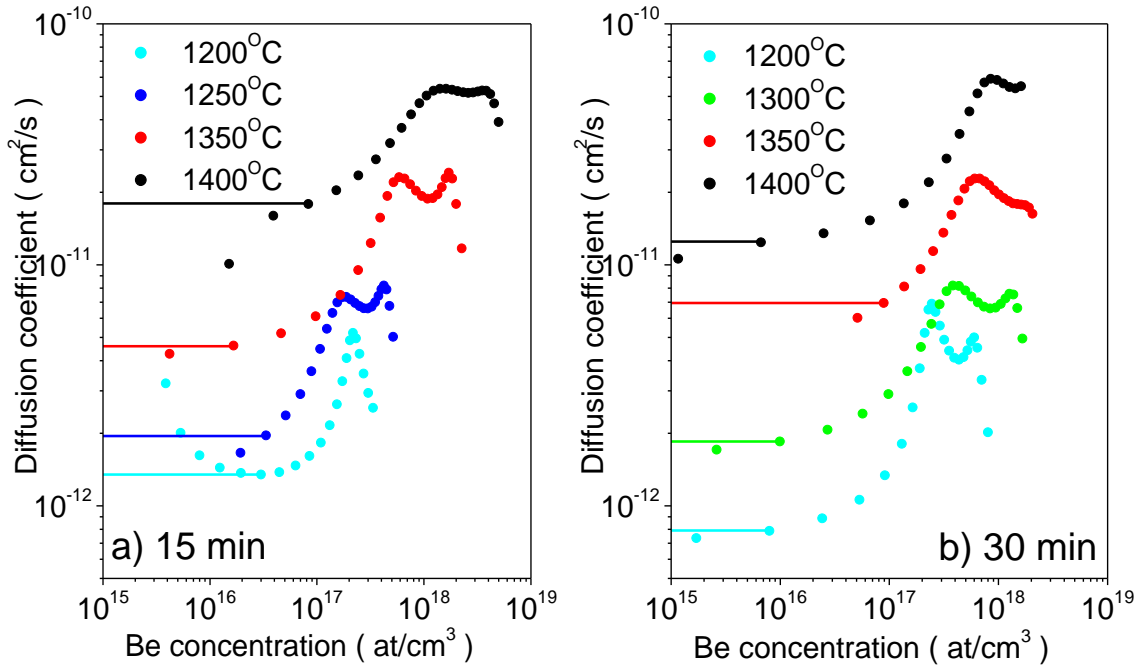

Fig. 3 Boltzmann-Matano analysis of Be depth profiles according to eq. (3) for samples annealed for: a) 15 minutes and b) 30 minutes; straight lines indicate diffusion coefficients for low concentration of Be; two peaks visible for high Be concentrations may be a consequence of the departure from the Boltzmann-Matano analysis at the initial Be profile.

The diffusion coefficients calculated from eq. (1) basing on the *erfc* fit and the coefficients derived from the Boltzmann-Matano analysis for both annealing times are presented as a function of 1000/T in Fig. 4. A classical Arrhenius equation can be used to fit the dependence of the diffusion coefficients on the inverse temperature:

$$D = D_O \exp\left(\frac{-E_A}{kT}\right) \qquad (4)$$

The result of this fit for data from the *erfc* and Boltzmann-Matano analysis are also presented in Fig. 4.

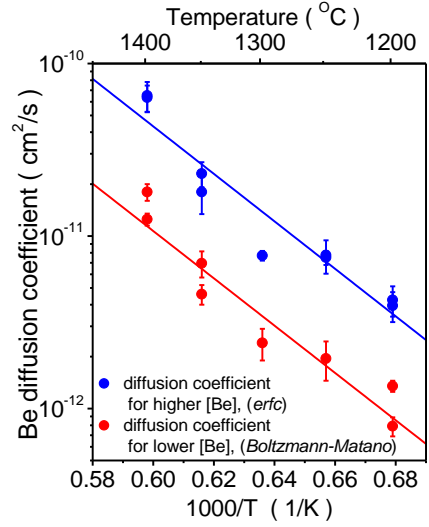

Fig. 4 Diffusion coefficients of Be atoms in GaN as a function of inverse temperature; Arrhenius plot from *erfc* fitting (blue filled circles and line) and Boltzmann-Matano analysis (red filled circles and line).

The fit from eq. (2) allowed to determine the temperature-independent pre-exponent factor $D_0$ and the activation energy for both Be diffusion mechanisms using the *erfc* fitting as well as the Boltzmann-Matano analysis. The results are presented in Table 1.

Table 1 Temperature-independent pre-exponent factor $D_0$ and the activation energy for the Be diffusion

|  | Pre-exponent factor $D_0$ (cm$^2$/s) | Activation energy (eV) |
| --- | --- | --- |
| Higher Be concentration (*erfc* fitting) | $7.8 \pm 1 \times 10^{-3}$ | $2.73 \pm 0.05$ |
| Lower Be concentration (Boltzmann-Matano analysis) | $1.8 \pm 1 \times 10^{-3}$ | $2.72 \pm 0.05$ |

## 4. Discussion

As shown in Fig.1 most of the implantation damage created by Be ions in HVPE-GaN was removed by the UHPA process. At high temperature diffusion mechanisms start in GaN and implantation damage could be removed. Thus, the diffusion of Be atoms in a high quality, unintentionally doped HVPE-GaN layers was examined.

In the standard case of diffusion from an infinite source, the diffused element is deposited on the sample's surface and the host material initially contains no diffused species. In the case of implanted samples, they do not contain the examined species before implantation. In as-implanted HVPE-GaN layers an initial profile of Be existed. As was shown in Fig. 1b the area around 1 μm from the surface can be treated as an infinite source of Be dopant for annealing conditions. In turn, as indicated in Fig. 1b, it took nearly 6 μm for a smooth decrease of Be concentration from about $10^{21}$ at/cm$^3$ to the SIMS detection limit of $10^{15}$ at/cm$^3$.

The Be profiles for samples annealed at different temperatures (see Figs 1b and 2) all exhibit a visible drop (kink). Similar behavior was already observed for diffusion of Zn in GaAs [22], Zn in InP [23], and Mn in GaAs [24]. As shown, the change in the SIMS profile shape indicates that the diffusion coefficient depends on the concentration of the studied element. In case of the measured Be profiles, the *erfc* function was fitted to the data in the upper part of the profiles presented in Fig. 2. This allowed to determine the diffusion coefficients for higher Be concentration. For lower Be concentration these coefficients were calculated from the Boltzmann-Matano method. Figure 3 shows the results of concentration-dependent diffusion coefficients obtained by this method. The coefficients at high Be concentration are in very good agreement with those derived from the *erfc* function. Below the kink, the diffusion coefficients are almost one order of magnitude lower. It should be noted that the activation energies determined by two described above approximations were the same, about 2.7 eV. The difference was only found in the value of temperature-independent pre-exponent factor $D_0$. Its values were greater for higher Be concentration (see Tab. 1). All these results may suggest two mechanisms, fast and slow, of Be diffusion in GaN.

It should be noted that the tails of Be diffusion go much deeper than the kink, which is possibly due to the diffusion pass through threading dislocations. In the studies which used heteroepitaxial layers with high threading dislocation density, this effect might be dominant. Therefore, the present research with the use of high-quality host material allows for a discussion of Be diffusion process by removing the effect of dislocations.

Literature data indicate Ga site as the most probable for Be location in GaN, when the Fermi level is closer to the conduction band minimum [6]. Since $V_{Ga}$ with concentration lower than

$10^{17}$ cm$^{-3}$ exist in undoped HVPE-GaN [25,26] and they can also be created during ion implantation process, Be atoms will most likely occupy the free Ga sites. It should be noted that the concentration level of the Be profile drop (black dash lines in Fig. 3) increased with the annealing temperature. Thermodynamics of defects in crystalline materials states that the concentration of defects increases with the temperature according to the formula:

$$C_d = N \exp\left(\frac{-E_f}{kT}\right) \quad (4)$$

where: $C_d$ is the defect concentration, $N$ is the number of potential sites in the lattice (per unit volume) where the defect can be created ($4.4 \times 10^{22}$ cm$^{-3}$ for Ga substitutional site in GaN), $E_f$ is the formation energy (energy needed to create a defect), $k$ is the Boltzmann constant, and $T$ is the temperature.

Gallium vacancies seem to be responsible for decelerating the diffusion of Be in the studied samples. The diffusion mechanisms depend on the local concentrations of Be atoms and $V_{Ga}$. When Be concentration is lower than that of $V_{Ga}$, Be atoms are trapped in Ga positions and the slower interstitial-substitutional diffusion mechanism associated with $V_{Ga}$ or $Be_{Ga}$-$Be_i$ complex diffusion dominates. Such conditions dominate at the Be diffusion front. Similar diffusion mechanism was shown for Si [27] or Mn [28] in GaN. When Be concentration exceeds the $V_{Ga}$ content, the excess amount of the dopant diffuses via the fast interstitial mechanism. Such switching of Be lattice site was previously shown by positron annihilation method [9].

The slower diffusion reveals a 4 times smaller pre-exponential factor which is proportional to the path of the atom in a single hop and the probability of finding a site to jump. Octahedral and tetrahedral positions are the most favorable for a Be atom in hexagonal GaN. Therefore, a pure interstitial diffusion path runs through the main channel of the hexagonal structure, i.e., along the *c* axis and via octahedral sites (see Fig 5a). In the case of interstitial-substitutional mechanism, two possible paths exist. They both involve a $V_{Ga}$ and either an octahedral (Fig. 5b) or tetrahedral position (Fig. 5c).

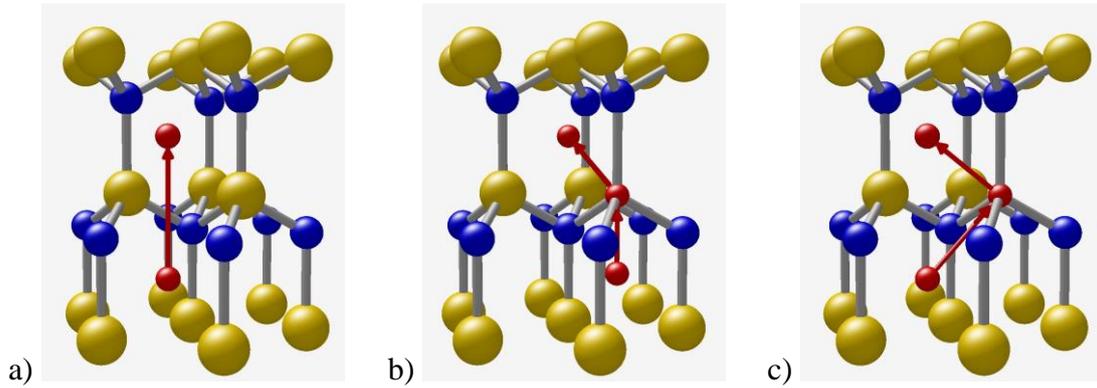

Fig. 5 Scheme of the Be diffusion path: a) interstitial diffusion via octahedral position, b) interstitial-substitutional diffusion via tetrahedral position and Ga vacancy, c) interstitial-substitutional diffusion via tetrahedral position and Ga vacancy.

Gallium vacancy concentration was determined from Fig. 2 according to the concentration level of the Be profiles drop. An Arrhenius plot for $V_{Ga}$ concentration is presented in Fig. 6 (red circles). The defect formation energy obtained from a linear fit is equal to $1.7 \pm 0.05$ eV. This value corresponds well to the $V_{Ga}$ formation energy calculated from first principles for n-type GaN annealed under N-rich conditions [29].

Another parameter which could be derived from the Be depth profiles is the formation energy of the dopant associated with its solid solubility in GaN at a given temperature. In case of diffusion from an infinite source the solid solubility corresponds to the $C_S$ parameter (see eq. (1)), which is the maximum Be concentration for the diffused part of GaN layer. Values of $C_S$, resulting from the fitting function to the Be diffusion profile, are also presented in Fig. 6 (blue filled circles).

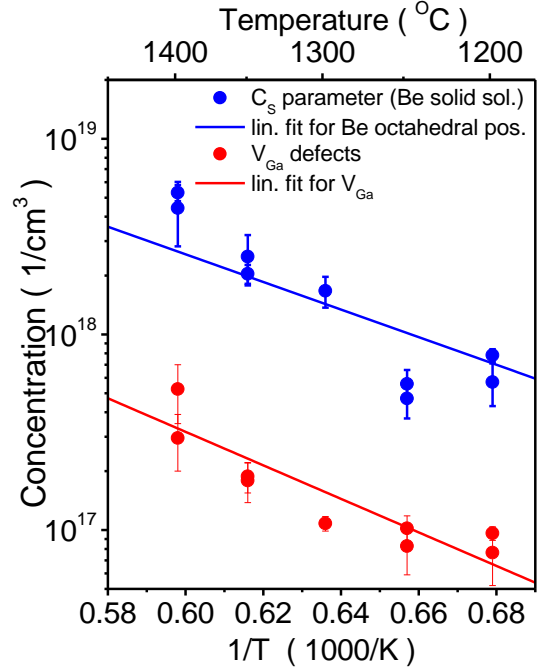

Fig. 6 Arrhenius plot for gallium vacancy $V_{Ga}$ concentration (red dots) and Be solid solubility ($C_S$ parameter, blue dots).

Taking into account the possible interstitial position of Be atoms, two most favorable positions mentioned before in the GaN hexagonal lattice are considered: octahedral (one site per cation) or tetrahedral (two sites per cation). For a proper fit of a linear function of Arrhenius plot the pre-exponential factor $N$ in eq (4) has to be stated as a constant. For one or two sites per Ga atom $N$ is equal to $4.4\times10^{22}$ cm$^{-3}$ or $8.8\times10^{22}$ cm$^{-3}$, respectively. As it was shown before, a rapid interstitial diffusion proceeds via octahedral positions, therefore the first value of $N$ should be stated. The fit performed for octahedral positions of Be, is presented in Fig. 6. The formation energy for an interstitial Be defect at octahedral position in hexagonal GaN is equal to $1.4 \pm 0.05$ eV. Based on the theoretical data concerning Be$^{2+}$ interstitial [6], the calculated formation energies indicate that after the Be diffusion the Fermi level in GaN crystal is situated in the middle of the bandgap. This is consistent with the experimentally observed (see Experimental setup) high resistivity of the implanted and annealed samples.

## 4. Conclusions.

In this study the mechanisms of Be diffusion in GaN were investigated and described. For the first time the temperature dependence of Be diffusion was established. It allowed to show concentration-dependent diffusion coefficient resulting from two different diffusion mechanisms. For both of them the activation energy was $E_A = 2.7$ eV. Different values of the pre-exponential factor, equal $D_O=7.8\times10^{-3}$ cm$^2$/s and $D_O=1.8 \times 10^{-3}$ cm$^2$/s, were determined for the faster and slower diffusion, respectively. The first process seems to be a pure interstitial mechanism through octahedral lattice sites. The slower one is an interstitial-substitutional diffusion mechanism involving Ga vacancies and tetrahedral lattice sites. The ratio of atoms involved in both mechanisms depends on the $V_{Ga}$ defect concentration. Such a result shows that by controlling the Ga vacancy concentration, the rate of diffusion of the Be dopant can be influenced. Further studies should be focused on the possibility of achieving a higher content of $V_{Ga}$ defects. It would allow to locate a larger amount of Be atoms in the Ga substitutional position and, thus, a greater or even total activation of dopant.

**Acknowledgment**

This research was supported by the Polish National Science Center through projects No. 2018/29/B/ST5/00338, as well as by TEAM TECH program of the Foundation for Polish Science co-financed by the European Union under the European Regional Development Fund (POIR.04.04.00-00-5CEB/17-00). The authors would like to thank Tetsuo Narita from Toyota Central R&D Labs for very detailed discussion regarding the diffusion processes in GaN, his support in the analysis of the presented results and helpful suggestions regarding the manuscript.

[1] H. Amano, Y. Baines, E. Beam, M. Borga, T. Bouchet, P. R. Chalker, M. Charles, K. J. Chen, N. Chowdhury, R. Chu Journal of Physics D: Applied Physics 51, 163001 (2018)

[2] Y. Irokawa, O. Fujishima, T. Kachi, and Y. Nakano, J. Appl. Phys. 97, 083505 (2005)

[3] Y. Nakano, T. Jimbo, Journal of Applied Physics 92, 3815 (2002)

[4] H. Sakurai, M. Omori, S. Yamada, Y. Furukawa, H. Suzuki, T. Narita, K. Kataoka, M. Horita, M. Boćkowski, J. Suda, and T. Kachi, Appl. Phys. Lett. 115 (14), 142104 (2019)

[5] T. Narita, H. Sakurai, M. Bockowski, K. Kataoka, J. Suda, T. Kachi, Applied Physics Express 12, 111005 (2019)

[6] J. L. Lyons, A. Janotti, C. G. Van de Walle, Jpn. J. Appl. Phys. 52, 08JJ04 (2013)

[7] F. J. Sanchez, F. Calle, M.A. Sanchez-Garcia, E. Calleja, E. Munoz, C.H. Molloyz, D.J. Somerford, J.J. Serrano and J.M. Blanco, Semicond. Sci. Technol. 13, 1130 (1998)

[8] Y. Nakano, T. Kachi, and T. Jimbo, Appl. Phys. Lett. 82, 2082 (2003)

[9] F. Tuomisto, V. Prozheeva, I. Makkonen, T. H. Myers, M. Bockowski, Henryk Teisseyre, Phys. Rev. Lett. 119, 196404 (2017)

[10] H.T. Wang, L.S. Tan, E.F. Chor, Journal of Crystal Growth 268, 489 (2004)

[11] C. Ronning, K. J. Linthicum, E. P. Carlson, P. J. Hartlie, D. B. Thomson, T. Gehrke, R. F. Davis, MRS Online Proceedings 537, G3.17 (1998)

[12] Y. Sun, L. S. Tan, S. J. Chua, S. Prakash, MRS Online Proceedings 595, F99W3.82 (1999)

[13] S. J. Chang, W. C. Lai, J. F. Chen, S. C. Chen, B. R. Huang, C. H. Liu, U. H. Liaw, Materials Characterization 49, 337 (2003)

[14] H. W. Huang, C. C. Kao, J. Y. Tsai, C. C. Yu, C. F. Chu, J. Y. Lee, S.Y. Kuo, C. F. Lin, H. C. Kuo, S.C. Wang, Materials Science and Engineering B 107, 237 (2004)

[15] O. Koskelo, U. Koster, F. Tuomisto, K. Helariutta, M. Sopanen, S. Suihkonen, O. Svensk and J. Raisanen, Phys. Scr. 88, 035603 (2013)

[16] G. Miceli, A. Pasquarello, Phys. Status Solidi RRL 11, 1700081 (2017)

[17] M. Zajac, R. Kucharski, K. Grabianska, A. Gwardys-Bak, A. Puchalski, D. Wasik, E. Litwin-Staszewska, R. Piotrzkowski, J. Z. Domagala, M. Bockowski, Prog. Cryst. Growth Charact. Mater. 64, Issue 3, 63 (2018)

[18] M. Bockowski, M. Iwinska, M. Amilusik, M. Fijalkowski, B. Lucznik, T. Sochacki, Semicond. Sci. Technol. 31, 093002 (2016)

[19] J. Karpiński, J. Jun, and S. Porowski, J. Cryst. Growth 66, 1 (1984).

[20] K. Lorenz, S.M.C. Miranda, E. Alves, I.S. Roqan, K.P. O'Donnell, M. Bockowski, Proceedings of SPIE 8262, 82620C (2012)

[21] C. Matano, Japan. J. Phys. 8, 109 (1932)

[22] L. R. Weisberg, J. Blanc, Physical Review 131, 1548 (1963)

[23] R. Jakiela, A. Barcz, E. Wegner, A. Zagojski, Vacuum 78, 417 (2005)

[24] R. Jakieła, A. Barcz, E. Wegner, A. Zagojski, Journal of Alloys and Compounds 423, 132 (2006)

[25] M. Bockowski, M. Iwinska, M. Amilusik, B. Lucznik, M. Fijalkowski, E. Litwin-Staszewska, R. Piotrzkowski, and T. Sochacki, J. Cryst. Growth 499, 1 (2018).

[26] R. Rounds, B. Sarkar, T. Sochacki, M. Bockowski, M. Imanishi, Y. Mori, R. Kirste, R. Collazo, and Zlatko Sitar, Journal of Applied Physics 124, 105106 (2018)

[27] R. Jakiela, A. Barcz, E. Dumiszewska, A. Jagoda, phys. stat. sol. (c) 3, 1416 (2006)

[28] R. Jakiela, K. Gas, M. Sawicki, A. Barcz, Journal of Alloys and Compounds 771, 215 (2019)

[29] JL Lyons and CG Van de Walle, npj Computational Materials 12 (2017)